\pdfoutput=1

\documentclass[12pt,a4paper,hyperref]{JHEP3}
\let\ifpdf\relax
\usepackage{ifpdf}
\usepackage{color}
\let\normalcolor\relax
\usepackage{amsmath}
\usepackage{graphics}
\usepackage{mathtools}
\usepackage{cite}
\newcommand{\bea}{\begin{eqnarray}}
\newcommand{\eea}{\end{eqnarray}}
\newcommand{\bean}{\begin{eqnarray*}}
\newcommand{\eean}{\end{eqnarray*}}
\newcommand{\nn}{\nonumber \\}

\renewcommand{\tilde}{\widetilde}
\newcommand\sss{\hspace{0.7pt}}
\newcommand{\stwo}[2]{s_{#1\sss #2}}

\renewcommand{\phi}{\varphi}
\newcommand{\chyInt}{d\Omega_{\mathrm{CHY}}}
\newcommand{\mi}{\raisebox{0.75pt}{\scalebox{0.75}{$\,-\,$}}}
\newcommand{\pl}{\raisebox{0.75pt}{\scalebox{0.75}{$\,+\,$}}}
\newcommand{\LR}{\raisebox{-2.25pt}{\,\scalebox{1.75}{$\Leftrightarrow$}}\,}

\newcommand{\fwbox}[2]{\text{\makebox[#1][c]{$\hspace{-150pt}\displaystyle#2\hspace{-150pt}$}}}

\newcommand{\eq}[1]{\vspace{-0.pt}\begin{equation}\hspace{-100pt}#1\hspace{-100pt}\vspace{-0.pt}\end{equation}}
\newcommand{\eqs}[1]{\vspace{-0.0pt}\begin{equation}\begin{split}#1\end{split}\vspace{-0.0pt}\end{equation}}
\newcommand{\fig}[3]{\raisebox{#1}{\includegraphics[scale=#2]{#3}}}

\newcommand{\z}[2]{(z_{#1}\!-\!z_{#2})}

\renewcommand{\phi}{\varphi}
\DeclareMathOperator*{\Res}{\mathrm{Res}}
\newcommand{\mell}{\!\mi\ell}

\preprint{2015}
\title{{\LARGE \mbox{Integration Rules for Loop Scattering Equations}}}
\author{{\normalsize \mbox{Christian~Baadsgaard$^1$, N.~E.~J.~Bjerrum-Bohr$^1$, Jacob~L.~Bourjaily$^1$,} \mbox{Poul~H.~Damgaard$^1$ and Bo~Feng\mbox{$^{1,2}$}}}\\
\mbox{{\mbox{$^1$}\ Niels Bohr International Academy and Discovery Center}}\\
\mbox{{\ \;\! The Niels Bohr Institute, University of Copenhagen}}\\
\mbox{{\ \;\! Blegdamsvej 17, DK-2100 Copenhagen, Denmark}}\medskip \\
\mbox{{\mbox{$^2$}\ Zhejiang Institute of Modern Physics}}\\
\mbox{{\ \;\! Zhejiang University}}\\
\mbox{{\ \;\! PR China, 310027}}}
\date{\today}
\abstract{We formulate new integration rules for one-loop scattering equations analogous to those at tree-level,
and test them in a number of non-trivial cases for amplitudes in scalar $\phi^3$-theory. This formalism greatly facilitates the evaluation of amplitudes in the CHY representation at one-loop order, without the need to explicitly sum over the solutions to the loop-level scattering equations.}

\begin{document}

\newpage
\section{Introduction}\label{sec:introduction}

The formalism of Cachazo, He and Yuan (CHY)~\cite{Cachazo:2013gna,Cachazo:2013hca,Cachazo:2013iea,Cachazo:2014xea} is an intriguing reformulation of quantum field theory that represents scattering amplitudes as integrals over an auxiliary coordinate space completely localized by $\delta$-functions which impose a set of algebraic constraints referred to as the {\it scattering equations}. At tree-level, the auxiliary integral is performed over points $z_i\!\in\!\mathbb{P}^1$ associated with each particle, and the scattering equations (which fully localize the $z_i$'s) correspond to
\vspace{0.5pt}\eq{S_i\equiv\sum_{i\neq j}\frac{\stwo{i}{j}}{\z{i}{j}}=0\label{scattering_equations}\,,\vspace{-2.5pt}}
for the $i^{\mathrm{th}}$ particle, with $\stwo{i}{j}\!\equiv\!(k_i+k_j)^2$ being the familiar Mandelstam invariants. The precise measure of integration for scattering amplitudes depends on the theory in question, but the constraints $\delta(S_i)$ always localize the integral to a sum over isolated solutions to the scattering equations (\ref{scattering_equations}). For $n$ particles, there are $(n\mi3)!$ solutions to these equations. Integration measures for many theories are known, and a proof of this remarkable construction for scalar $\phi^3$-theory and Yang-Mills theory has been given by Dolan and Goddard in \mbox{ref.\ \cite{Dolan:2013isa}.}

In practice, the summation over $(n\mi3)!$ solutions makes the formalism very cumbersome already at rather low multiplicity
kinematics. Recently, two complementary methods were developed that circumvent this brute-force procedure and which directly
produce the result of integration---that is, summing over all the solutions~\cite{Cachazo:2015nwa,Baadsgaard:2015voa}. Moreover, a direct link
between individual Feynman diagrams and integrands for the CHY representation has been
provided as well~\cite{Baadsgaard:2015ifa}. With this, one has complete control over the CHY construction at tree-level and is
therefore ready to tackle the question of amplitudes at loop-level.

There are two obvious paths towards obtaining a scattering equation formalism valid at loop-level. With the now known map
between CHY-integrands and tree-level Feynman diagrams, one could make use of generalized unitarity to reconstruct loop
amplitudes out of on-shell, tree-level diagrams and use the tree-level scattering equations. A more elegant solution would
build on the close connection between the CHY-formalism and string theory~\cite{Mason:2013sva,Berkovits:2013xba,Gomez:2013wza,Bjerrum-Bohr:2014qwa}.
Indeed, steps in that direction were taken in \mbox{ref.~\cite{Adamo:2013tsa}} and further developed in \mbox{ref.~\cite{Casali:2014hfa}},
identifying field theory loops in terms of the genus expansion, as in string theory. The main, na\"{i}ve stumbling block in that approach
is the natural appearance of elliptic functions that, in ordinary perturbation theory, should be represented as integrals over rational
functions. A breakthrough in this direction has recently been made by Geyer, Mason, Monteiro and Tourkine~\cite{Geyer:2015bja}.
In the context of supergravity, they show how to reduce the problem of genus one to a modified problem on the Riemann sphere,
where the analysis is essentially the same as at tree-level. They provide a conjecture for the $n$-point supergravity one-loop amplitude,
and suggest how to generalize their result to any loop-order; they also provide a conjecture for super Yang-Mills amplitudes at one-loop.

In this paper, we generalize the analysis of \mbox{ref.~\cite{Geyer:2015bja}}, and show how it naturally leads to a representation of one-loop amplitudes in $\phi^3$-theory. The scalar case provides the simplest setting in which to understand the use of
scattering equations at loop-level. As discussed in \mbox{refs.~\cite{Casali:2014hfa,Geyer:2015bja}}, the one-loop case essentially
amounts to computing an $n$-point amplitude by means of an auxiliary $(n\pl 2)$-point scattering amplitude involving two additional particles with momenta $\ell$ and $\mell$ (that is, taken in the forward limit). Intuitively, this is not unlike representing loops using the Feynman tree theorem \cite{Feynman:1963ax,Brandhuber:2005kd}, for example. However, the representation of amplitudes using the scattering equations appears quite a bit more magical as we will see below.

An essential ingredient that makes the scattering equation formalism work at loop-level is the freedom to shift what becomes loop momentum $\ell$ by an arbitrary constant in any individual term---a
property that must be respected by the regularization framework being used.\footnote{This is the case for dimensional regularization. Because the scattering equation formalism is independent of the number of spacetime dimensions, it is natural for us to use it here. See also the discussion in section 4.} This is because, as we will see, the scattering equation formalism naturally generates rather unfamiliar representations of loop integrands---involving `propagators' that are almost exclusively linear in the loop momentum.

The loop-level scattering equations are nearly identical to those at tree-level, but with two additional particles with opposite (off-shell) momenta. As such, there are $(n\pl2\mi3)!\!=\!(n\mi1)!$ solutions in general. This counting differs from that of \mbox{ref.\ \cite{Casali:2014hfa}} because we use loop-level scattering equations that differ due to regularization concerns that will be discussed in \mbox{section \ref{scalar_phi3_section}}. And we will find that the integration rules described in \mbox{ref.~\cite{Baadsgaard:2015voa}} must be modified slightly to take into account the additional, off-shell momenta in the forward limit. The principal difference will be that for $\phi^3$-theory, our representation explicitly removes tadpole contributions (similar to the dimensionally-regulated Feynman expansion). Although this paper is mainly concerned with scalar $\phi^3$-theory, it is clear that the integration rules we describe can be applied to a much broader class of theories.

Our paper is organized as follows. In the next section we provide a lightning review of the scattering equation formalism,
including the integration rules that permit us to evaluate terms without the explicit summation over solutions to
the scattering equations. In \mbox{section \ref{one_loop_section}} we turn to loop-level, using the recent supergravity solution of
\mbox{ref.~\cite{Geyer:2015bja}} as a guide for inferring the correct integration measure for scalar $\phi^3$-theory.
We test this proposal in \mbox{section \ref{scalar_phi3_section}} with concrete examples at one-loop.

\section{Scattering equations and integration rules at tree-level}\label{review_and_tree_section}

Recall that in the CHY formalism, ordered tree-level scattering amplitudes in massless $\phi^3$-theory can be represented \cite{Cachazo:2013hca,Dolan:2013isa} as follows:
\eq{\mathcal{A}_n^{(\phi^3),\text{tree}} = \int\!\chyInt\left(\frac{1}{\z{1}{2}^2\z{2}{3}^2\!\cdots\z{n}{1}^2}\right)\,.\label{n_point_function_in_phi3_thy}}
Here, $\chyInt$ represents a universal integration measure together with the $\delta$-function constraints which impose scattering equations (\ref{scattering_equations}) (and fully localize the integral):
\eq{\!\!\!\!\!\!\!\!\!\!\!\!\!\!\!\chyInt\equiv\frac{d^nz}{\mathrm{vol}(SL(2,\mathbb{C})\!)}
\prod_i\,\!'\delta(S_i)=\!\z{r}{s}^2\z{s}{t}^2\z{t}{r}^2
\prod_{\fwbox{25pt}{i\!\in\!\mathbb{Z}_n\!\backslash\{r,s,t\}}} dz_i\,\delta(S_i)\,.\label{definition_of_chy_measure}}
This measure is independent of the $SL(2,\mathbb{C})$ gauge-choice of points labelled $\{r,s,t\}$. Because the $\delta$-functions fully localize the integral (\ref{n_point_function_in_phi3_thy}), it becomes simply a sum over the $(n\mi3)!$ isolated solutions to the scattering equations.

Scattering amplitudes in different theories can all be represented as integrals over $\chyInt$, but with different integrands than that of (\ref{n_point_function_in_phi3_thy}). More generally then, we will be interested in integrals of the form:
\vspace{2.5pt}\eq{\int\!\chyInt\,\,\,\mathcal{I}(z_1,\ldots,z_n)\,.\label{general_chy_int}}
For the sake of concreteness, let us restrict our attention to M\"{o}bius-invariant integrals involving products of factors of the form $\z{i}{j}$ (with $i\!<\!j$) in the denominator. We can represent integrands of this form graphically by drawing vertices for each $z_i$, and connecting vertices $\{z_i,z_j\}$ for each factor of $\z{i}{j}$ appearing in the denominator. M\"{o}bius-invariance requires that  each factor $z_i$ occurs four times, resulting in integrands represented by four-regular graphs. For example, consider the integrand represented graphically by,
\vspace{-7.5pt}\eq{\hspace{-10pt}\fig{-36.375pt}{1}{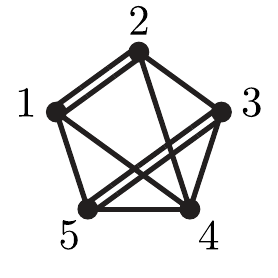}\hspace{-10pt}\LR\frac{1}{\z{1}{2}^2\z{2}{3}\z{3}{4}\z{4}{5}\z{1}{5}\z{3}{5}^2\z{1}{4}\z{2}{4}}.\nonumber\vspace{-7.5pt}}
Integration of this function $\mathcal{I}(z_1,\ldots,z_5)$ against the measure $\chyInt$ results in an inverse product of Mandelstam invariants---in this case, $1/(\stwo{1}{2}\stwo{3}{5})$.

A combinatorial rule for the result of integration for integrals of the form (\ref{general_chy_int}) was described in \mbox{ref.\ \cite{Baadsgaard:2015voa}}, which we briefly summarize here. Integrals of this form generally result in a sum of inverse-products of multi-index Mandelstam invariants denoted $s_{i \:\! j \:\! \cdots \:\! k}\!\equiv\!s_{\{i,j,\ldots,k\}}\!\equiv\!(k_i\pl k_j\pl\cdots\pl k_k)^2$ (for arbitrary subsets $P\!\subset\!\{1,\ldots,n\}$). In general, each term in the sum will be a product of precisely $(n\mi3)$ factors,
\vspace{-5pt}\eq{\prod_{a=1}^{n-3}1/s_{P_a},\vspace{-5pt}\label{tree-pole}}
where each $P_a\!\subset\!\{1,\ldots,n\}$ denotes a subset of legs that we can always take to have at most $n/2$ elements (because $s_{P}\!\!=\!s_{P^\complement}$, with $P^\complement\!\equiv\!\mathbb{Z}_n\backslash P$, by momentum conservation). The collections of subsets $\{P_a\}$ appearing in (\ref{tree-pole}) must satisfy the following criteria:\\[-24pt]
\begin{itemize}
\item for each pair of indices $\{i,j\}\!\subset\!P_a$ in each subset $P_a$, there are exactly $(2|P_a|\mi2)$ factors of $\z{i}{j}$ appearing in the denominator of $\mathcal{I}(z_1,\ldots,z_n)$;\\[-24pt]
\item each pair of subsets $\{P_a,P_b\}$ in the collection is either nested or complementary---that is, $P_a\!\subset\!P_b$ or $P_b\!\subset\!P_a$ or $P_a\!\subset\!P_b^\complement$ or $P_b^\complement\!\subset\!P_a$;\\[-24pt]\end{itemize}
if there are no collections of $(n\mi3)$ subsets $\{P_a\}$ satisfying the criteria above, the result of integration will be zero.

These integration rules produce the result of the integration in eqn.~(\ref{n_point_function_in_phi3_thy}) for an arbitrary number of external legs in tree-level $\phi^3$-theory. In the next section, we will need integration rules for loop integrands of one-loop with $(n\pl2)$ external legs, two of which are neighboring with off-shell momenta $\ell$ and $\mell$. The rules will be quite similar to those described above, but with a few small changes. One prominent change will be the appearance of Mandelstam-like objects generalized to include off-shell momenta:
\eq{[i,j,\ldots,k]\equiv(k_i+ k_j+\cdots+ k_k)^2-(k_i^2+ k_j^2+\cdots+ k_k^2)\,.\label{cycle-def}}
Notice that $[i,j,\ldots,k]$ becomes identical to $s_{i \:\! j \:\! \cdots \:\! k}$ when all the momenta are on-shell and massless.

\section{Scattering equations for one-loop amplitudes}\label{one_loop_section}

The scattering equations at one-loop-level given in \mbox{ref.~\cite{Geyer:2015bja}} provide a great simplification
over the ones considered in \mbox{refs.~\cite{Adamo:2013tsa,cdgTalks}}. We refer to those references for details.

At tree-level, the scattering equations are defined on the Riemann surface as discussed above.
The locations of the external legs are parametrized by the coordinates, $z_i$, where $i$ runs from $1$ to $n$ for the $n$-point amplitude.
At one-loop level one has to consider scattering equations on the torus---the genus-one surface.
Here, $\tau$ and $z$ parametrize the torus, and the points $z_i$ has the same meaning as
in the tree-level case, {\it i.e.}, they are the positions of the external legs.
At one-loop the scattering equations are
\bea \Res_{z_i} P(z,z_i|q)^2= 2k_i\cdot P(z=z_i,z_i|q)=0\,,~~~~~P(z=z_0,z_i|q)^2(z_0)= 0\,, \eea
where $z_0$ is an arbitrary point on the torus and the one-form $P(z,z_i|q)$ is the solution to the following differential equation
\bea
\overline{\partial} P(z,z_i|q)=2\pi i\sum^n_i k_i \overline \delta(z-z_i) dz\,.
\eea
The solution can be parametrized by
\bea P(z,z_i|q)= 2\pi i \ell dz +\sum_{i}^n k_i \left( {\theta_1'(z-z_i)\over \theta_1(z-z_i)}
+\sum_{j\neq i} {\theta_1'(z_{ij})\over n\theta_1(z_{ij})}\right) dz\,,\eea
on the torus where $q$ is related to the modular variable $\tau$ in the following
way: $q=e^{2\pi i\tau}$. $\ell$ will
turn out to play the role of the loop momentum. $\theta_1(z)$ is the
standard modular function that also appears in string theory.

The one-form $P(z,z_i|q)$ can be greatly simplified in the limit $q=e^{2\pi i\tau}\to 0$, where $\tau\to +i\,\infty$, and
by changing variables from $z_i$ to $\sigma_i$  and $z$ to $\sigma$ using the following redefinitions: $\sigma_i=e^{2\pi i(z_i-\tau/2)}$, $\sigma=e^{2\pi i(z-\tau/2)}$.
In the new variables translational invariance of $z$ becomes scaling invariance of $\sigma$, ({\it i.e.} $dz={d\sigma\over 2\pi i\sigma}$), and in the limit one observes that
\bea {\theta_1'(z-z_i)\over \theta_1(z-z_i)} dz\to {-d\sigma\over 2\sigma}+{d\sigma\over \sigma-\sigma_i}\,.\eea
Using momentum conservation $(\sum_i^n k_i){-d\sigma\over 2\sigma}=0 $ in the limit yields
\bea P(z,z_i|q)\to P(\sigma,\sigma_i)= \ell {d\sigma\over \sigma}+ \sum_{i}^n { k_i d\sigma\over \sigma-\sigma_i}\,,\label{Pz-exp}
\eea
after redefining  $\ell \to \ell-\sum^n_{i< j} (k_i-k_j) \cot(\pi z_{ij})  {1\over 2 i n}$.
We now find that
\bea\hspace{-0.5cm}  P(\sigma,\sigma_i)^2 -{\ell^2\,d\sigma^2\over \sigma^2} & = &
\sum_i^n {2\ell\cdot k_i \,d\sigma^2 \over \sigma(\sigma-\sigma_i)}
+\sum^n_{ i<j} {2k_i \cdot k_j \, d\sigma^2\over (\sigma-\sigma_i)(\sigma-\sigma_j)}\,.
\label{P2z-exp}
\eea
The combination $P(\sigma,\sigma_i)^2-{\ell^2 d\sigma^2\over \sigma^2}$ has only single poles.
It is easy to calculate the residues of these single poles and they are
\bea {\cal S}_i\equiv {[\ell,k_i] \over \sigma_i}
+\sum^n_{j\neq i} {[i,j]\over (\sigma_i-\sigma_j)}\,,~~~~\label{SC-loop-ki}
\eea
for the single pole at $\sigma_i$ and
\bea {\cal S}_{0}\equiv \sum^n_i {[\ell,i]\over -\sigma_i}\,,~~~~\label{SC-loop-0}
\eea
for the single pole at $\sigma=0$. The residue of   $\sigma=\infty$ is zero. It is easy to check that
$\sum_{i=1}^n {\cal S}_i=-{\cal S}_0$. Furthermore,  $\sum_{i=1}^n \sigma_i {\cal S}_i=0$.
The equations defined by ${\cal S}_{0} = 0$ and ${\cal S}_i=0$ are the one-loop scattering
proposed in~\cite{Geyer:2015bja} on the Riemann sphere, with  $\ell$ playing the role of the loop momentum.
As shown above only $(n\mi1)$ of these equations are independent.
If we compare them with the tree scattering equations, it is clear that the one-loop scattering equations for $n$-point
amplitudes are very similar to the tree-level scattering equations for $(n\pl2)$ external legs, where two legs of off-shell momenta $\ell,\mell$ have been inserted and
fixed to the values $\sigma_{\ell}=0$ and $\sigma_{-\ell}=\infty$. To avoid confusion we will distinguish 
the tree-level case from the one-loop case by using $z_i$ for the insertions at tree-level and $\sigma_i$ for the insertions at one-loop level.
One crucial difference between the tree-level case and the one-loop case is that we take $\ell$ and $\mell$ to be off-shell.

Since two points $0,\infty$ have been fixed ($\sigma_{\ell}=0$ and $\sigma_{-\ell}=\infty$), the general $SL(2,\mathbb{C})$-transformation
on the Riemann sphere ${a\sigma+b\over c\sigma+d}$ is reduced to just ${a\sigma\over d}$. This means that we are in the
one-loop case just left with a scaling invariance, which, using $ad-bc=1$  reads $\sigma\to a^2\sigma$.  The scaling invariance can be
immediately observed in the scattering equations~(\ref{SC-loop-ki}) and can also be understood from the definition
$\sigma=e^{2\pi i(z-\tau/2)}$. The scaling symmetry in the $\sigma_i$ coordinates corresponds to
translational invariance in the original one-loop torus variables.

Our goal now is to find the correct CHY measure at loop-level for color ordered $\phi^3$ theory, insisting on the scaling invariance discuss above.
We will start the discussion by recalling the tree-level measure \medskip\medskip
\eq{\!\!\!\!\!\!\!\!\!\!\!\!\!\!\!\chyInt\equiv\frac{d^nz}{\mathrm{vol}(SL(2,\mathbb{C})\!)}
\prod_i\,\!'\delta(S_i)=\!\z{r}{s}^2\z{s}{t}^2\z{t}{r}^2
\prod_{\fwbox{25pt}{i\!\in\!\mathbb{Z}_n\!\backslash\{r,s,t\}}} dz_i\,\delta(S_i)\,.~~~
}
Introducing $z_{ij}\!\equiv\!\z{i}{j}$, we can write tree-level amplitudes in the following general form
\bea \int\left(\prod_{i=1}^n d z_i\right)\left(z_{rs}z_{st}z_{tr} \prod_{a\neq r,s,t}
\delta\left( {S}_a\right)\right)\left( {1\over {\cal F}(z)}\right)\left( {1\over d\omega}\right)\,.~~~~\label{tree-m-1}
\eea
Now let us analyze the four factors in~(\ref{tree-m-1}). Since we have only $(n\mi3)$ independent scattering equations, we correspondingly insert
only $(n\mi3)$ $\delta$-function constraints. However, the result must be independent of the choice of which equations we choose.
This independence is precisely achieved by the factor $z_{rs}z_{st}z_{tr}$ that is inserted in the measure and which renders the combined
expression permutation invariant. This factor provides also the same
transformation under the $SL(2,\mathbb{C})$ group as that of the three scattering equations that have been removed. Because of these
 first two factors in eqn.~(\ref{tree-m-1}), ${\cal F}$ must transform as
\eq{{\cal F}(z)\to \left(\prod_{i=1}^n {(ad-bc)^2\over (c z_i+d)^4} \right){\cal F}(z)\,,}
under the $SL(2,\mathbb{C})$ transformation
\eq{z_i \to \frac{az_i + b}{cz_i +d}\,.}
Different choices of this factor ${\cal F}$ with proper transformation properties will define different theories. The last factor
$d\omega\!\equiv\!{dz_r dz_s dz_t\over z_{rs}z_{st}z_{tr}}$ provides the Koba-Nielsen gauge fixing.

Having understood how the integrand is composed for a tree-level amplitude in the CHY formalism, we now proceed to deduce
the corresponding integrand at one-loop level.
First, since there are now only $(n\mi1)$ independent loop scattering equations, we can have only $(n\mi1)$ $\delta$-function
constraints $\delta({\cal S}_i)$. Again, to make the result independent of the choice of which equation we eliminate, we
need to insert a factor with the same scaling property as the $\delta$-function we removed. A natural combination is
$\left(\sigma_l\prod_{j\neq l}^n \delta({\cal S}_j)\right)$.~\footnote{The same choice can also be inferred
from the corresponding factor at tree-level: The term $z_{ij}z_{jk}z_{ki}$ with $z_{i=\ell}=0$ and $z_{j=-\ell}=\infty$ reduces to $z_{ki}=z_k$.}
Now (in a similar way to the tree-level case) we can write down the proposed integration at one-loop level
\bea \int   {1\over {\rm vol}(GL(1))}  \left(\prod_{i=1}^n d\sigma_i \right) \left(\sigma_l\prod_{j\neq l}^n \delta({\cal S}_j)\right)
\left( {1\over {\cal F}(\sigma_i)}\right)\,. ~~~~\label{1loop-mu}
\eea
Scaling invariance now requires that ${\cal F}(\lambda \sigma_l)=\lambda^{2n} {\cal F}(\sigma_l)$. Using the standard Faddeev-Popov
method, we can gauge fix any $\sigma_k$ to a fixed value. We will call this the $(k,l)$ gauge-choice, where $l$ is the scattering equations
removed and $k$ is the $\sigma_k$ that has been fixed. With this gauge choice eqn.~(\ref{1loop-mu}) reads
\bea \int    \left(\prod_{i=1}^n d\sigma_i \right) \left(\sigma_l\prod_{j\neq l}^n \delta({\cal S}_j)\right)\left( {1\over {\cal F}(\sigma_i)}
\right)\left( {1\over d\omega}\right),~~~~d\omega={d\sigma_k\over \sigma_k}\,. ~~~~\label{1loop-mu-1}
\eea
Next we will consider the possible choices of ${\cal F}(\sigma_i)$ corresponding to different theories, such as gravity, Yang-Mills theory,
and scalar field theory at one-loop level.

For gravity there is no color ordering, the amplitude must be symmetric in the external legs and we therefore require that ${\cal F}(\sigma_i)$
is totally permutation invariant. The scaling degree $2n$ leads to the natural choice ${\cal F}(\sigma_i)={\cal I}^{-1}G^2$ with
$G=\prod_{i=1}^n \sigma_i$ and ${\cal I}^{-1}$ being a scale invariant expression. An example for $\cal I$ in supergravity has been
conjectured in \mbox{ref.~\cite{Geyer:2015bja}} with the gauge fix $(k,l)=(1,1)$.

For Yang-Mills theory, \mbox{ref.~\cite{Geyer:2015bja}} conjectured the following factor to go into the expression for ${\cal F}(\sigma_i)$
\bea PT_n(\gamma) & = & {\sigma_{\ell(-\ell)}\over \sigma_{\ell\gamma(1)}\sigma_{\gamma(1),\gamma(2)}\ldots
\sigma_{\gamma(n-1),\gamma(n)}\sigma_{\gamma(n)(-\ell)}}\,,~~~\label{1loop-PT}\eea
where $\gamma$ is an element of the $n$-point permutation group $\mathfrak{S}_n$. Noting that we have $(n\pl1)$ factors in denominator
and only one in numerator, the scaling degree of $PT$ is $n$. To arrive at the overall scaling of degree $2n$ we thus need another
factor in ${\cal F}(\sigma_i)$ with scaling degree $n$. It is natural to assume that the other factor is $G$, defined above. Thus,
for a given color ordering $\gamma$ we should expect ${\cal F}_\gamma(\sigma_i)={\cal I}^{-1}PT_n(\gamma)\, G$
where ${\cal I}^{-1}$ again is a scale invariant expression. After taking the gauge fixing $(k,l)=(1,1)$, we arrive at
the expression in \mbox{ref.~\cite{Geyer:2015bja}}. A possible $\cal I$ for super Yang-Mills theory has been conjectured in \mbox{ref.~\cite{Geyer:2015bja}.}

Now we will concern ourselves with the scalar case. Having gained experience from the supergravity and super
Yang-Mills theory cases, it is natural to assume that for color ordered bi-adjoint scalar $\phi^3$-theory, we should
have ${\cal F} (\sigma_i)=PT_n(\gamma_1) PT_n(\gamma_2)$ with $\gamma_1,\gamma_2$ being two
permutations in $\mathfrak{S}_n$. In other words, the na\"ive expectation would be for the scalar amplitude $\cal A$ to be given by
\bea {\cal A}(\gamma_1|\gamma_2)\equiv \int   {\prod_{i=1}^n d\sigma_i \over d\sigma_k}\sigma_l\sigma_k\prod_{j\neq l}^n \delta({\cal S}_j) PT_n(\gamma_1) PT_n(\gamma_2)\,. ~~~~\label{mab}
\eea
The analogue quantity of ${\cal A}(\gamma_1|\gamma_2)$ at tree-level is ${m}(\gamma_1|\gamma_2)$ in \mbox{ref.~\cite{Cachazo:2013iea},}
which is nothing but the  inverse of the momentum kernel ${\cal S}[\gamma_1|\gamma_2]$ that was first defined
in~\cite{BjerrumBohr:2010ta,BjerrumBohr:2010yc,BjerrumBohr:2010hn}.
We thus have ${\cal S}[\gamma_1|\gamma_2]={m}(\gamma_1|\gamma_2)^{-1} $, with \eq{{\cal S}[i_1,\ldots,i_k | j_1,\ldots, j_k] = \prod_{t=1}^k\left(s_{{i_t} 1}+\sum_{q>t}^k \Theta(i_t,i_q)s_{{i_t},{i_q}}\right),} where $\Theta$ is the Heaviside function. The function $\cal A$ at loop-level can be thought of as the
inverse (one-loop) momentum kernel.

However it turns out that the na\"ive choice for $\cal A$ above is not yet complete. Using the mapping between the
integrand and the CHY graph, we see that unlike the tree-level case where we have a closed CHY graph, $PT$ is an open chain
connecting the two fixed nodes $\sigma_{\ell}=0$ and $\sigma_{-\ell}=\infty$ through the $n$ external points $\gamma(1),\ldots,\gamma(n)$.
It is crucial that the integrand does not contain {\it an inverse
factor of $\sigma_{\ell,({-\ell})}$}. Instead, we in fact have to insert a numerator factor of
$\sigma_{\ell,({-\ell})}$ so that the integrand becomes scale invariant.

The appearance of an open chain is related to the physical picture that after cutting a one-loop propagator the closed loop is
opened up, but the additional two legs are not physical external states. Since we can cut any loop propagator, this
physical picture also suggests that to get the complete one-loop integrand of a given color ordering, we should sum
over all cyclic orderings. In other words, the pair $\{\ell,\mell\}$ should be inserted at all possible places
of the given color ordering of $n$-points. From this we are now led to the correct compact expression:
\bea {\cal A}_{\phi^3}(\gamma)\equiv (-1)^n\int {d^d\ell\over \ell^2}\int   {\prod_{i=1}^n d\sigma_i \over d\sigma_k}\sigma_l
\sigma_k\prod_{j\neq l}^n \delta({\cal S}_j) \sum_{\text{cyclic}} (PT_n(\gamma))^2\,,  ~~~~\label{oneloop-phi3}\eea

Having obtained this proposal~(\ref{oneloop-phi3}) for one-loop scalar amplitudes, we now use the $\delta$-function
constraints to integrate out the $\sigma_i$'s. Using~(\ref{SC-loop-ki}), it is straightforward to find the elements of the Jacobian,
\bea {\partial{\cal S}_i\over \partial\sigma_j} & = & {[i,j]\over (\sigma_i-\sigma_j)^2},~~~~i\neq j\,, \nn
{\partial{\cal S}_i\over \partial\sigma_i} & = & -{[\ell,i] \over \sigma_i^2}
-\sum_{j\neq i} {[i,j]\over (\sigma_i-\sigma_j)^2}\,.~~~~~\label{Si-Jacobi}\eea
Furthermore with the choice of $\sigma_{\mell}=\infty$, the $PT$ factor is simplified to
\bea PT_n(\gamma) & = & {1\over \sigma_{\ell\gamma(1)}\sigma_{\gamma(1)\gamma(2)}\ldots\sigma_{\gamma(n-1)\gamma(n)}}\,.~~~\label{1loop-PT-a-fix}
\eea
Putting all these pieces together, we finally arrive at
\bea
{\cal A}_{\phi^3}(\gamma)= (-1)^n\int {d^d\ell\over \ell^2} \sum_{\text{cyclic}}\sum_{\text{solutions}} {\sigma_l \sigma_k
\over (-)^{l+k}{\cal J}({\cal S})_l^k}(PT_n(\gamma))^2\,,~~~~\label{1loop-measure-2}\eea
where the ${\cal J}({\cal S})_l^k$ is the determinant of Jacobian matrix after deleting the $l$-th row and $k$-th column,
and the sum runs over the solutions to the loop-level scattering equations. Although there is also a sum over cyclic
permutations of $\gamma$ in eqn.~(\ref{1loop-measure-2}), we need to calculate only one set, obtaining the others trivially by relabeling.

\begin{figure}
 \centering\fig{-36.375pt}{1}{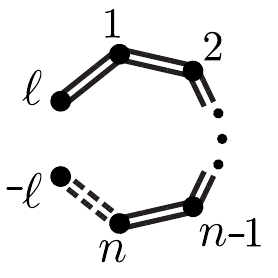}
  \caption{The CHY graph for $n$-points with ordering $\{1,2,\ldots,n\}$. The dashed line between $n,-\ell$ will disappear
  when we impose the gauge fixing $\sigma_{-\ell}=\infty$.  }\label{nn}
\end{figure}

Just as at tree-level, we can associate a CHY graph with the one-loop integrand $(PT_n(\gamma))^2$ in~\eqref{1loop-measure-2}.
Such a one-loop graph for the integrand is illustrated in Fig. \ref{nn} with the Koba-Nielsen gauge fixing $\sigma_{\mell}\to \infty$.
For such graphs, we can immediately use the integration rules of \mbox{ref.~\cite{Baadsgaard:2015voa}} with two minor modifications.
The final result can still be presented in the form of eqn.~(\ref{tree-pole}) which will provide the full result of the integration in~(\ref{1loop-measure-2}) without explicitly solving the one-loop scattering
equations and summing over all of them. The two modifications are the following. First, instead of having poles ${1\over s_{P_a}}$,
we must replace them by ${1\over [P]}$ where the notation $[P]$ has been defined by eqn.~\eqref{cycle-def}. In the massless
case, the two expressions are the same, but for off-shell momenta with $\ell^2 \neq 0$, they are different.
Secondly, we should explicitly exclude the set $P=\{\ell,\mell\}$ (or its
complement)\footnote{Obviously, a set $P$ with only one element (or its complement) should not be included, neither at
tree-level nor at the one-loop level.}. Not including the set  $P=\{\ell,\mell\}$ eliminates diagrams with singular zero-momentum
propagators associated with tadpoles. As a side remark we would like to note that it is also possible to write up the specific
individual Feynman diagrams at loop-level; such a decomposition will be similar to a $n$-gon decomposition into triangle
diagrams as was considered in \mbox{ref.~\cite{Baadsgaard:2015ifa}.}

\section{Scalar one-loop amplitude examples}\label{scalar_phi3_section}

In this section, we will demonstrate that the results obtained by solving the
one-loop scattering equations using the integration measure proposed above match those obtained from the
Feynman diagram expansion at one-loop order, after the proper
regularization of the singular terms associated with zero momentum propagation. Furthermore, these results can be
obtained directly from the associated loop-level CHY graph using our loop-level integration rules. 

We will start with the one-loop integrand for the two-point `amplitude' of $\phi^3$-theory. Although this example
is quite singular, it is simple enough to demonstrate many features of our calculation. In particular, the augmented
four-point amplitude with two additional external legs $\ell$ and $-\ell$ is well defined and is in fact the simplest
example to start with. We will first present the calculation in terms of Feynman diagrams, then explicitly use the
scattering equations, and finally present the corresponding CHY graph and the result of employing the loop integration rule.
\\
\\
\noindent{\bf Using Feynman diagrams:} Without considering the tadpole diagram, there is only one
term in the one-loop integrand,
\bea {1\over \ell^2 (\ell+k_1)^2}\,,\eea
corresponding to the diagram

\begin{figure}[h]
  \centering\fig{-13.85pt}{1}{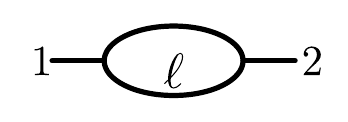}
  \caption{The Feynman diagram at two points.}\label{Feynman_2p_bubble}
\end{figure}

Using the general partial fraction formula
\bea {1\over \prod_{i=1}^n D_i} & = & \sum_{i=1}^n {1\over D_i \prod_{j\neq i} (D_j-D_i)}\,,~~~\label{PF}
\eea
that was also exploited in \mbox{ref.~\cite{Geyer:2015bja}}, we can split the integrand into
\bea {1\over \ell^2 (2\ell\cdot k_1)} +{1\over  (\ell+k_1)^2 (-2\ell\cdot k_1)}= {1\over \ell^2 (2\ell\cdot k_1)} +{1\over
\tilde\ell^2 (2\tilde\ell\cdot k_2)}\,,~~~\label{2p-bubble-split}
\eea
where we have used the on-shell condition $k_1^2=0$ and defined the variable $\tilde\ell=\ell+k_1$ for
the second term. Since, with a proper regularization (such as dimensional regularization), we can freely shift the
loop momentum,
we can identify $\tilde\ell=\ell$ in the second term of~(\ref{2p-bubble-split}) and write
\bea {1\over \ell^2  [\ell,1]}+ {1\over \ell^2  [\ell,2]}\,.~~~\label{2p-bubble-Fe}
\eea
In fact, using that $k_2=-k_1$ we now see that the sum in~\eqref{2p-bubble-Fe}, the integrand of the on-shell bubble diagram, adds up to zero. This assumes that the integration really has been properly regularized so that the shift is allowed. Around $d=4$ dimensions the massless $\phi^3$ theory we are considering suffers from both ultraviolet and infrared divergences. Also, a mass term is not protected, and is thus expected to be generated in this theory at loop level from precisely this kind of two-point function: the infrared divergences already give a strong hint that such a mass generation will occur. Indeed, in this theory a massless on-shell particle can decay into two in the forward direction by the self-interaction, thus making the very definition of the $S$-matrix of the exactly massless theory subtle at the quantum level \cite{Dashen:1974dv}. It is probably best to consider the massless theory only around $d=6$ dimensions, where it is classically scale invariant and perturbatively renormalizable.

Let us also emphasize some points about the result~\eqref{2p-bubble-Fe}. First,
 the two terms are related to each other by $\mathbb{Z}_2$ cyclic permutation. As we will see, this is a general feature.
 Secondly, although they sum up to be zero, each term will appear in different
 orderings $PT_n(\gamma)$ when we use  the scattering equations. Thus it is necessary to write them in the form shown in~\eqref{2p-bubble-Fe}. A similar phenomenon occurs in all later examples.
\\
\\
\noindent{\bf Using the one-loop scattering equations:} To use the  setup presented  in the previous section, we
need to make a gauge choice $(k,l)$, {\it i.e.}
choose which scattering equation ${\cal S}_l$ is to be removed and which variable $\sigma_k$ is to be fixed.
However, when we do this in this two-point example (a highly singular case), a subtle point appears. The reason is the following.
After using momentum conservation, the two scattering equations become (keeping $k_1^2\neq 0$ as regulator
at the intermediate level of calculations)
\bea {\cal S}_1={[\ell,1]\over \sigma_1}+{[1,2]\over \sigma_1-\sigma_2}=0\,,~~~~~~{\cal S}_2={[\ell,2]
\over \sigma_2}+{[1,2]\over \sigma_2-\sigma_1}=0\,.
\eea
This leads to the identity ${[\ell,1]\over \sigma_1}={[\ell,1]\over \sigma_2}$.
Thus for general $\ell\cdot k_1\neq 0$, we arrive at $\sigma_1=\sigma_2$. In other words,
we cannot gauge fix $\sigma_1=1$ and leave $\sigma_2$ to be a free variable.
Thus we have to introduce another type of regulator $\mu$:
\bea {\cal S}_1={[\ell,1]\over \sigma_1}+{\mu\over \sigma_1-\sigma_2}=0\,,~~~~~~{\cal S}_2=-{[\ell,1]
\over \sigma_2}-{\mu\over \sigma_2-\sigma_1}=0\,.
\eea

Because of the special (singular) kinematics associated with the pair $\{\ell,\mell\}$ that introduces on-shell bubbles (we denote bubbles on-shell or off-shell depending on the nature of their external legs),
to arrive at well defined results, we need to sum over cyclic orderings
before we remove the regularization.

Choosing the color ordering $\gamma=\{1,2\}$ and taking the gauge choice $(k,l)=(1,1)$, we get for the integrand
\bea {-1\over \mu}+{-1\over -\mu+[\ell,2]}\,,\eea
Similarly the same gauge choice for the color ordering $\gamma=\{2,1\}$ will lead to
\bea  {1\over \mu}+{1\over [\ell,2]}\,.\eea
We see that adding these two terms together and carefully taking the limit $\mu\to0$,
we get again a zero result as in eqn.~\eqref{2p-bubble-Fe}.

\begin{figure}
 \centering\fig{-36.375pt}{1}{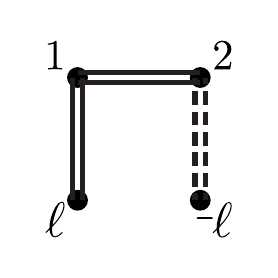}\quad\fig{-36.375pt}{1}{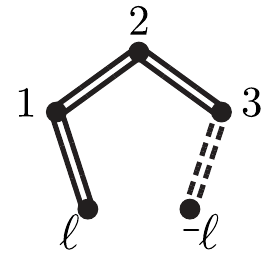}\quad\fig{-36.375pt}{1}{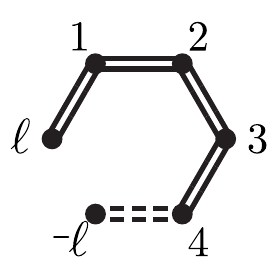}
  \caption{The CHY graphs for two-point (left), three-point  (middle) and four-point(right).
    }\label{2p3p}
\end{figure}

~\\{\bf Interpretation via a CHY graph:} We now present the corresponding CHY graph given by $PT(\gamma)^2$. For the ordering  $\gamma=\{1,2\}$, the graph is the following: we have four ordered
nodes $\{ \ell, 1,2, \mell\}$, and their connections are $\{ (\ell,1)_2, (1,2)_2\}$. Here we have used subscript to indicate
how many lines connect two nodes (see \mbox{Figure~\ref{2p3p})}. Using the tree mapping rule, na\"ively, we get
following possible poles:
${1\over[\ell,1]}$, ${1\over[1,2]}$. However, the complement of the pole ${1\over[1,2]}$ is ${1\over[\mell,\ell]}$ which has been removed
explicitly in the definition of CHY diagram ({\it i.e.}, there is no such denominator in the integrand
 $(PT_2)^2$), so we should not include it. This is the modification of the integration rule we need when it is applied at one-loop level.
Thus we are left with only
the pole $1/[\ell,1]$, which gives the final expression ${1\over \ell^2[\ell,1]}$. Including the other cyclic permutation,
we end up with the same result as using the scattering equations.

Having done the two point example, we will next move on to the next simplest thing, the one-loop integrand of the
color ordered three-point amplitude.

~\\~\\{\bf Using Feynman diagrams:} For the  color-ordered integrand of amplitude $A(1,2,3)$, there is one triangle
and three on-shell bubbles related by $\mathbb{Z}_3$ cyclic symmetry\footnote{Again we will not include the tadpole diagrams.}. The triangle is  given by
\bea T_{3;(1|2|3)}& =& {1\over \ell^2 (\ell+k_1)^2(\ell-k_3)^2}= {1\over (2\ell\cdot k_1)(-2\ell\cdot k_3) \ell^2}\nn
& & + {1\over (-2\ell\cdot k_1)( -2\ell\cdot k_1-2\ell\cdot k_3) (\ell+k_{1})^2} +{1\over (2\ell\cdot k_3)(2\ell\cdot k_1+2\ell\cdot k_3) (\ell-k_3)^2}\nn
& = & {1\over \ell^2[\ell,1][3,\mell]}+{1\over \ell^2[\ell,2][1,\mell]}+{1\over \ell^2[\ell,3][2,\mell]}\,, ~~~~\label{3p-Fe-tri}
\eea
where from the second to the third equation, we have used a shift of momentum $\ell$, which of course is valid only under the
integration. It is easy to see that these three terms are related by
$\mathbb{Z}_3$ cyclic permutations.
\begin{figure}
  \centering\fig{-25.6pt}{1}{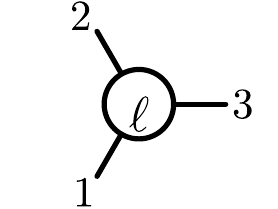}
  \caption{The triangle contribution at three points.}\label{triangle123}
\end{figure}
Similarly we can split  the three on-shell bubbles that are related by cyclic ordering.
\begin{figure}
  \centering\fig{-26.5pt}{1}{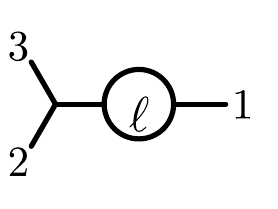}\qquad\fig{-26.5pt}{1}{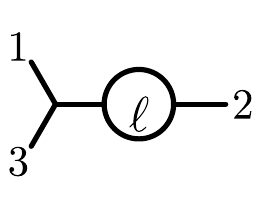}\qquad\fig{-26.5pt}{1}{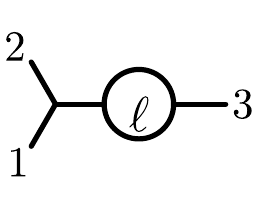}
  \caption{The three bubble contributions at three points.}\label{12_3_bubble}
\end{figure}
A typical one is\footnote{For an on-shell amplitude, we will have $s_{23}=0$. Thus to have a well defined meaning, one
should regularize $k_i^2\neq 0$ for the legs $i=1,2,3$.  }
\bea T_{2;(1|23)}&= & {1\over \ell^2 (\ell+k_1)^2 s_{23}}={1\over \ell^2[\ell,1][2,3]}+ {1\over \ell^2[2,3][1,\mell]}\,. ~~~~\label{3p-Fe-bub}
\eea

To compare with the results from scattering equations and CHY graphs, we reorganize all $1\times 3+3\times 2=9$
terms into three groups, which are related to each other by $\mathbb{Z}_3$ cyclic permutations. The first group is
\bea {\cal G}^{(3p)}_1 & = & {1\over \ell^2[\ell,1][3,\mell]}|_{T_{3;(1|2|3)}}+ {1\over \ell^2[\ell,1][2,3]}|_{T_{2;(1|23)}}+ {1\over \ell^2[1,2][3,\mell]}|_{T_{2;(3|12)}}\,,~~~\label{3p-G1}
\eea
where we have used the subscript to indicate where this term comes from. In fact, as we will see, ${\cal G}^{(3p)}_1$
is given by the CHY graph with ordering  $\gamma=\{1,2,3\}$. Again summing over three cyclic permutations, the
on-shell bubble part cancels and we are left with only the triangle contribution.

~\\{\bf Using the scattering equations:} We now use the scattering equations to find the integrand.
Let us start with ordering $\gamma=\{1,2,3\}$. As expected, one will get contributions from the on-shell
bubbles $(1|2+3)$ as well as $(1\pl2|3)$. To regulate the solutions we
set $k_1^2\neq 0$ and $k_3^2\neq 0$.
For the gauge choice $(k,l)=(1,1)$ we get
\bea &&  {\mell\!\cdot\! k_2\over 4(k_1\!\cdot\! k_2) (\mell\!\cdot\! k_1+k_1\!\cdot\! k_3)(\ell \!\cdot\! k_3-k_3^2)}\nn
\!\!\!\!\! &\!\!\!\!\!\!\! = & {1\over 4(k_1\!\cdot\! k_2)}\!\!
\left( {1\over (\mell\!\cdot\! k_1+k_1\!\cdot\! k_3)}+{-1\over (\ell \!\cdot\! k_3-k_3^2)}
+{(k_1\!\cdot\! k_2)\over (\mell\!\cdot\! k_1+k_1\!\cdot\! k_3)(\ell \!\cdot\! k_3-k_3^2)}\right).~~~~\label{3p-i1}
\eea
Taking  the limit of $k_1^2, k_3^2\to 0$ we get
\bea {1\over [\ell,1][2,3]} +{1\over [1,2][3,\mell]}
+{1\over [\ell,1][3,\mell]}\,,~~~\label{3p-SE}
\eea
which, when inserting the $1/\ell^2$-factor, is the same as ${\cal G}_1^{(3p)}$ in~\eqref{3p-G1}.

In the three point case having done the ordering $\gamma=\{1,2,3\}$, we should add the other two orderings $\gamma=\{3,1,2\}$ and $\gamma=\{2,3,1\}$ related by cyclic permutations. Summing all three contributions we match the Feynman expansion independently
of the gauge.

~\\{\bf Interpretation via a CHY graph:} We now present the corresponding CHY graph derivation given by the
integrand with  $\gamma=\{1,2,3\}$: with the ordering of nodes $\{ \ell,1,2,3,\mell\}$, thus the
connections are $\{ (\ell,1)_2, (1,2)_2, (2,3)_2\}$ (see \mbox{Figure \ref{2p3p})}. Using the mapping rule, we have
the following possible poles (again, since ${1\over[1,2,3]}={1\over[\mell,\ell]}$ we do not include these poles)
\bea & & {1\over[\ell,1]},~~~{1\over[1,2]},~~~{1\over[2,3]},~~~{1\over[\ell,1,2]}\,. \eea
Taking the compatible combinations we get the following result for the propagators
\bea {1\over[\ell,1][2,3]},~~~~~{1\over[\ell,1,2][\ell,1]},~~~~{1\over[\ell,1,2][1,2]}\,.\eea
Thus we have exactly the contribution ${\cal G}^{(3p)}_1$. Again
adding the cyclic permutations we arrive at the complete answer.

The four-point amplitude is the first non-trivial example where we can really test the formalism. Again we will employ three different paths to get the result, and compare them.

~\\ {\bf Using Feynman diagrams:} We first write down the color ordered one-loop integrand using Feynman diagrams. There is one box diagram
\eqs{\hspace{-40pt}T_{4;(1|2|3|4)}&= {1\over \ell^2 (\ell+k_1)^2 (\ell+k_{12})^2 (\ell-k_4)^2}\\
&= \phantom{+\,}{1\over \ell^2 [\ell,1][\ell,1,2][4,\mell]}+{1\over \ell^2[\ell,2][4,1,\mell][1,\mell]}\\
&\phantom{=\,}+ \,{1\over \ell^2[\ell,3][1,2,\mell][2,\mell]}
+{1\over \ell^2 [\ell,4][\ell,4,1][3,\mell]}\,.\label{4p-Fe-box}}
Here we have used a momentum shift to reach the last line.
\begin{figure}[h]
  \centering\fig{-36.375pt}{1}{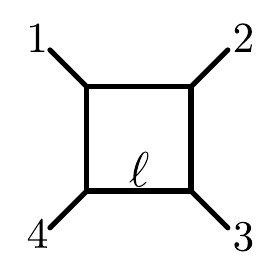}
  \caption{The box contribution $T_{4;(1|2|3|4)}$ at four points.}\label{1234box}
\end{figure}
Using  identities such as $[\ell,1,2]=[3,4,\mell]$ for four-point kinematics, it is easy to see that these
four terms in~\eqref{4p-Fe-box} are  related by $\mathbb{Z}_4$ cyclic permutations.
Next there are four triangles  related to each other by a $\mathbb{Z}_4$ cyclic permutation. As an example, we can consider the triangle contribution
\begin{figure}[h]
  \centering\fig{-26.5pt}{1}{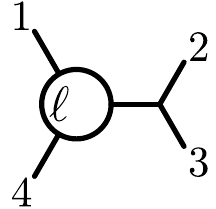}
    \caption{One of triangle contributions $T_{3;(1|23|4)}$ at four points.}\label{23_triangle}
\end{figure}
\eqs{\hspace{-30pt}T_{3;(1|23|4)} &=   {1\over \ell^2 (\ell+k_1)^2 (\ell-k_4)^2 s_{23}},\\
&={1\over \ell^2[\ell,1][2,3][4,\mell]}+{1\over \ell^2[2,3][4,1,\mell][1,\mell]}+{1\over \ell^2[\ell,4][\ell,4,1][2,3]}\,.\hspace{-50pt}~~~\label{4p-Fe-T31}}
For the bubbles, there are two different kinds in this four-point case: off-shell bubbles and on-shell bubbles.
For the on-shell bubbles, there are four which are related by a $\mathbb{Z}_4$ cyclic permutation. The first one is (again we use the intermediate
regularization $k_i^2\neq 0$ to make them well-defined)
\begin{figure}[h]
  \centering\fig{-26.5pt}{1}{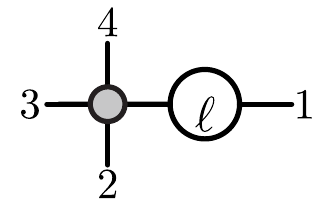}%
  \caption{The bubble contribution $T_{2;(1|234)}$ at four points.}\label{234_1_bub}
\end{figure}
\eqs{ \hspace{-60pt}T_{2;(1|234)} &=  {1\over \ell^2 (\ell+k_1)^2 s_{34} s_{234}}\\
&=\phantom{+\,}{1\over \ell^2 [\ell,1][2,3,4][3,4]}+{1\over \ell^2 [2,3,4][3,4][1,\mell]}\\
&\phantom{=\,}+{1\over \ell^2 [\ell,1][2,3][2,3,4]}+{1\over \ell^2 [2,3][2,3,4][1,\mell]}\,.}
There are two off-shell bubbles. They are related by a $\mathbb{Z}_2$ permutations ({\it i.e.}, $1\to 2, 2\to 3, 3\to 4, 4\to 1$).  %
\begin{figure}[h]
  \centering\fig{-26.5pt}{1}{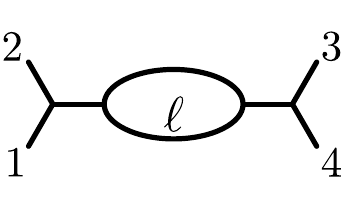}
  \caption{The bubble contribution $T_{2;(12|34)}$ at four points.}\label{12_34_bub}
\end{figure}
The first one is
\bea & & \hspace{-0.65cm}T_{2;(12|34)}  =  {1\over s_{12}\ell^2 (\ell+k_{12})^2 s_{34}}= {1\over \ell^2[\ell,1,2] [1,2][3,4]}+ {1\over \ell^2 [3,4] [1,2] [1,2,\mell]}\,.\eea
Now we reorganize all $36$ terms to  four groups, which are all related to each other by $\mathbb{Z}_4$ cyclic
permutations. The first group is
\bea & & {\cal G}^{(4p)}_1 = \nn & & {1\over \ell^2 [\ell,1][\ell,1,2][4,\mell]}|_{T_{4;(1|2|3|4)}}+{1
\over \ell^2[\ell,1][2,3][4,\mell]}|_{T_{3;(1|23|4)}}+ {1\over \ell^2[\ell,1][\ell,1,2][3,4]}|_{T_{3;(2|34|1)}}\nn
& & +{1\over \ell^2[1,2][3,4,\mell][4,\mell]}|_{T_{3;(4|12|3)}}+ \left(  {1\over \ell^2 [\ell,1][2,3,4][3,4]}+{1
\over \ell^2 [\ell,1][2,3][2,3,4]}\right)|_{T_{2;(1|234)}}\nn
& & + \left({1\over \ell^2 [1,2,3][2,3][4,\mell]}+{1\over \ell^2 [1,2][1,2,3][4,\mell]} \right)|_{T_{2;(4|123)}}\nn && +{1\over  \ell^2[\ell,1,2] [1,2][3,4]}|_{T_{2;(12|34)}}\,,~~~\label{4p-Fe-G1}\eea
where we have used the subscript to indicate where each contribution comes from.

~\\{\bf Interpretation via a CHY graph:} We now use the CHY graph procedure to reproduce the result from
the Feynman diagram expansion. Again, we need to sum up four graphs related by $\mathbb{Z}_4$ cyclic permutation.
The first one will be the graph with ordering $\{\ell,1,2,3,4,\mell\}$ and the connections
$\{ (\ell,1)_2, (1,2)_2, (2,3)_2, (3,4)_2\}$ defined by corresponding $PT$-factor (see \mbox{Figure \ref{2p3p})}. We list all possible poles:
\bea \text{double-pole:}  & ~~ &  {1\over[\ell,1]}\,,~~~{1\over[1,2]}\,,~~{1\over[2,3]}\,,~~~{1\over[3,4]}\,, \nn
\text{triple-pole:} & ~~ &  {1\over[\ell,1,2]}={1\over[3,4,\mell]}\,,~~~{1\over[1,2,3]}\,,~~~{1\over[2,3,4]}\,, \nn
\text{quadruple-pole:} & ~~ & {1\over[\ell,1,2,3]}={1\over[4,\mell]}\,.\eea
This yields various combinations of compatible propagators. There are five combinations containing two $2$-leg poles:
\bea && {1\over[\ell,1][2,3][4,\mell]}\,,~~{1\over[\ell,1][2,3][2,3,4]}\,,~~{1\over[\ell,1][\ell,1,2][3,4]}\,,\nn &&
{1\over[\ell,1][2,3,4][3,4]}\,, ~~{1\over[\ell,1,2][1,2][3,4]}\,.\eea
There are four combinations containing only a single $2$-leg-pole:\\[-14pt]
\bea {1\over[\ell,1][3,4,\mell][4,\mell]}\,,~~{1\over[\ell,1,2][1,2][4,\mell]}\,,\nn {1\over[1,2][1,2,3][4,\mell]}\,,~~ {1\over[1,2,3][2,3][4,\mell]}\,.\  \eea
These nine terms correspond exactly to the nine terms in ${\cal G}^{(4p)}_1$~\eqref{4p-Fe-G1}.
~~\\ ~\\ {\bf Using scattering equations:} Finally, we need to produce ${\cal G}^{(4p)}_1$ using the scattering equations
under the ordering $\gamma=\{1,2,3,4\}$. Again, to have well-defined results, we
regularize it with $k_1^2\neq 0, k_4^2\neq 0$. As one can check, there are
six solutions (in general $(n\mi1)!$ solutions for $n$-point). A numerical check yields the result  ${\cal G}^{(4p)}_1$ using the gauge fix $(k,l)=(4,4)$.
\\
\\
Before we end this section, let us briefly discuss the number of contributions generated by CHY graphs and by
Feynman diagrams. We will show that the counting with the new one-loop rules is still one-to-one, just as in the tree-level case.

For a given CHY graph, the combinations of compatible propagators that we count up are exactly those that appear in the color ordered tree-level $(n\pl2)$-point amplitude with extra legs $l$ and $-l$, except that we exclude the subset $\{l,-l\}$, which corresponds to removing all Feynman diagrams associated with $\ell,\mell$ attached to the same vertex.
These Feynman diagrams correspond to the color ordered tree-level amplitude with $(n\pl1)$-points. Thus using
the known formula for the number of color ordered diagrams in $\phi^3$ theory  with  $n$  external legs
\bea  C_n={2^{n-2}(2n\mi 5)!!\over (n\mi1)!}\,,\eea
we know immediately that each CHY  graph will give $C_{n+2}-C_{n+1}={3(n\mi1)2^{n-1}(2n\mi3)!!\over (n\pl1)!}$ terms. When summing
over cyclic orderings, we get a total number of
\bea {\cal T}_{CHY}(n)={3n(n\mi1)2^{n-1}(2n\mi3)!!\over (n\pl1)!}\,.~~~\label{terms}\eea

On the other hand  for each $n$-gon in the Feynman diagram expansion, after partial fractioning, we have $n$ terms,
corresponding to the $n$ choices of opening up  a single propagator. After each such opening-up of a propagator, we get a
color ordered tree-level Feynman diagram with $(n\pl2)$-points. Different openings give different orderings,
where the pair $\{\ell, \mell\}$ is inserted between different nearby vertexes $\{i,i+1\}$. Again, the Feynman
diagrams obtained this way do not contain pairs of $\ell,\mell$ attached to the same vertex. They are again
tree-level Feynman diagrams with $(n\pl1)$-points. Combining everything we get the counting
\bea  n( C_{n+2}-C_{n+1})\,,\eea
which is identical to the one given in eqn.~\eqref{terms}.

\section{Conclusion and discussion}

We have shown how the diagrammatic integration rules for scattering equations that were first
developed for tree-level amplitudes have an immediate
extension to one-loop level.  The integration rules at loop
level follow from those at tree-level with the following modification: the loop CHY integrand
has to be compensated so that it scales correctly. This naturally leads to valid integrands for the different
kinds of theories. Here we have spelled out in great detail how the procedure does appear to produce
correct integrands for scalar $\phi^3$-theory by systematically working through the
low-point cases. When considering scattering equations at loop-level it is essential to specify a regularization, and for the
procedure to work we need to be able to shift loop momentum by constants in the integrand. A
regularization scheme such as dimensional regularization should ensure this.
Because we have only been interested in demonstrating the mechanism through which the scattering equation formalism at loop level can generate the correct set of diagrams, we have ignored all issues that arise when actually performing the loop integration. In particular, the propagators should of course be given the usual $i\epsilon$-prescription of Feynman propagators.
 
The procedure that we have presented seems to be generalizable to higher loops. At each loop
order two more legs are added at the intermediate step. This is one obvious extension to
pursue in the future.

\subsection*{Acknowledgements}
We thank Freddy Cachazo, David Gross, Henrik Johansson, Arthur Lipstein, Lionel Mason and Ricardo Monteiro for discussions.
This work has been supported in part by a MOBILEX research grant from the Danish Council for
Independent Research (JLB), by the National Science Foundation under Grant No. NSF PHY11-25915 (PHD) and by Qiu-Shi funding and Chinese NSF funding under contracts No.11031005, No.11135006 and No.11125523 (BF). PHD and BF would like to thank for the hospitality at the KITP and NBIA, respectively.

\newpage

\providecommand{\href}[2]{#2}\begingroup\raggedright\endgroup

\end{document}